\documentclass[journal=jacsat,manuscript=article]{achemso}
\usepackage[version=3]{mhchem}
\usepackage{amsmath}
\usepackage{mathtools}
\usepackage{amsfonts}
\usepackage{amssymb}
\usepackage{diagbox}
\usepackage{dcolumn} 
\usepackage{array}
\newcolumntype{P}[1]{>{\centering\arraybackslash}p{#1}}
\newcolumntype{M}[1]{>{\centering\arraybackslash}m{#1}}
\newcolumntype{C}[1]{>{\centering\arraybackslash}p{#1}}
\usepackage{color}
\usepackage{float}
\usepackage{psfrag}
\usepackage{graphicx}
\usepackage{caption}
\usepackage{tikz}
\usetikzlibrary{positioning}
\usetikzlibrary{arrows}
\usetikzlibrary{trees}
\usepackage{midfloat}
\usetikzlibrary{decorations.pathmorphing}
\usetikzlibrary{decorations.markings}
\usetikzlibrary{automata,positioning}
\usepackage{braket}
\usepackage{rotating}
\usepackage{subfig}
\usepackage{epsfig}
\usepackage{adjustbox}
\usepackage{ragged2e}
\usepackage{simplewick}
\usepackage{simpler-wick}
\usepackage{csquotes}
\usepackage{xcolor}
\usepackage{fancyhdr}
\usepackage{multirow}
\usepackage{csquotes}
\UseRawInputEncoding
\author{Valay Agarawal} \affiliation{Department of Chemistry, Indian Institute of Technology Bombay, \\ Powai, Mumbai 400076, India} 
\author{Samrendra Roy} \altaffiliation{Contributed equally to this work}\affiliation{Department of Energy Science and Engineering, Indian Institute of Technology Bombay, \\ Powai, Mumbai 400076, India} 
\author{Kapil K. Shrawankar}\altaffiliation{Contributed equally to this work} \affiliation{Department of Chemistry, Indian Institute of Technology Bombay, \\ Powai, Mumbai 400076, India} 
\author{Mayank Ghogale}\affiliation{Institute of Chemical Technology, \\ Mumbai 400019}
\author{S Bharathi} \affiliation{Department of Chemistry, Indian Institute of Technology Bombay, \\ Powai, Mumbai 400076, India} 
\author{Anchal Yadav} \affiliation
{Department of Chemistry, Indian Institute of Technology Bombay, \\ Powai, Mumbai 400076, India}
\author{Rahul Maitra}\email{rmaitra@chem.iitb.ac.in} 
 \affiliation{Department of Chemistry, Indian Institute of Technology Bombay, \\ Powai, Mumbai 400076, India}
\title{Assessing the Performance of Nonlinear Regression based Machine Learning Models to Solve Coupled Cluster Theory}
\begin{document}
\begin{abstract}
The iterative solution of the coupled cluster
equations exhibits a synergistic relationship among the various 
cluster amplitudes. The iteration scheme may be viewed as a
multivariate 
discrete time propagation of nonlinearly coupled equations, 
which is dictated by only a few principal
cluster amplitudes. These principal amplitudes usually
correspond to only a few valence excitations, whereas all
other cluster amplitudes are enslaved, and behave as
auxiliary variables. Staring with a few trial iterations, 
we employ a supervised machine learning strategy to
establish an interdependence between the principal and 
auxiliary amplitudes.
We introduce a coupled cluster - machine learning hybrid 
scheme where the coupled cluster equations are solved only
to determine the principal amplitudes, which saves 
significant computation time. The auxiliary amplitudes, 
on the other hand, are determined via regression. Few
different regression
techniques have been introduced to express the auxiliary
amplitudes as functions of the principal amplitudes. 
The scheme has been applied to several molecules in their
equilibrium and stretched geometries, and our scheme, 
with all the regression models, shows a significant reduction
in computation time over the canonical coupled cluster 
calculations without unduly sacrificing the accuracy.
\end{abstract}
\maketitle
\section{Introduction:}
\label{intro}
Coupled cluster theory \cite{cc3,cc4,cc5,bartlett2007coupled} is a well
established method for solving the electronic
Schr{\"o}dinger wave equation for small to medium sized
atoms and molecules.  
CC theory employs an exponential wave operator $\Omega$, 
that brings in the effects of the excited Slater 
determinants on to the reference ground state 
wavefunction, which is usually taken to be the
Hartree-Fock determinant. The wave operator $\Omega$ 
is chosen as 
$\Omega = e^T$, where $T$ is the sum of all possible
many-body hole to particle excitation operators. In the 
most common cases, $T$ consists of one and two-body
hole-particle excitation operators. The resultant theory, 
known as the CC with singles and doubles excitations 
(CCSD) predicts accurate quite energy for molecules with
predominantly single reference character. The amplitudes
corresponding to the excitation operators, which are the
unknown quantities, are determined by projecting the
similarity transformed effective Hamiltonian 
$G=e^{-T}He^T$ against
the excited state determinants. The correlated ground 
state energy is calculated as the expectation value of 
the effective Hamiltonian with respect to the chosen
reference function, $E_{corr}=\langle \phi_{HF} | H_{eff} | \phi_{HF}\rangle=\langle \phi_{HF} | e^{-T} H e^T | \phi_{HF} \rangle$. Due to the exponential nature of the
wave operator, the similarity transformed Hamiltonian, 
$H_{eff}$ is highly non-linear in $T$, and hence one 
employs iterative scheme to solve these equations. For 
CCSD, the most expensive computational step scales as 
$n_o^2n_v^4$ per iteration, where $n_o$ is the number of
the occupied orbitals, and $n_v$ is the number of the 
virtual orbitals in the chosen reference determinant. 
This often makes the theory prohibitively time consuming
for large systems. The theory since it's inception has seen 
many developments to increase it's accuracy with a reduced 
computational scaling. Due to the iterative nature of the 
solutions, there have been significant efforts to 
accelerate the convergence\cite{pulay1980convergence,piecuch1994solving,kjonstad2020accelerated,yang2020solving} or scale down the
steep computational scaling associated with the solution
scheme\cite{parrish2019rank,schutski2017tensor,deprince2013accuracy,schutz2003linear}. One may thus look for an iterative scheme where 
the expensive $n_o^2n_v^4$ scaling may be bypassed, at
least partially, which is likely to save a lot of
computation time. This may be achieved by exploiting the
nonlinearity of the iterative solution scheme. 
In line with this, some of the present authors imbibed
ideas from nonlinear dynamics and synergetics\cite{haken1982slaving,Haken1983,Haken_1989} to
demonstrate that not all the cluster amplitudes are
equally important in the nonlinear iteration scheme. 
Based on the magnitude of the amplitudes, the authors
classified the amplitudes into \enquote{unstable master 
amplitudes} (later to be referred to as the principal 
amplitudes) and \enquote{stable slave amplitudes} (later 
to be referred to as the auxiliary amplitudes). Borrowing 
the key concepts from nonlinear discrete time-series
\cite{marwan2007recurrence,RecurrencePlot}, coupled with
principles imported from the areas of Synergetics, a
Machine Learning (ML)\cite{murphy2012machine} based 
hybrid numerical scheme was developed to establish a 
relationship between the two classes of amplitudes. This 
effectively reduces the independent degrees of freedom
to accelerate the overall iteration process.
In the resulting scheme, only a few initial iterations scale as
$n_o^2n_v^4$, while most of the other iterations scale
significantly less. This saves a lot of computation time, 
as was demonstrated by the authors. Complementary to that,
one may employ an adiabatic approximations based on the
difference in the time scale of relaxation of various 
cluster amplitudes during the iteration, where one may
formally reduce the scaling of the iterative scheme by at 
least one order of magnitude, without undue sacrifice of
the accuracy. 

In section II, we discuss the essential aspects of the
newly developed iteration scheme, where we reiterate the 
existence of a circularly causal relationship among the 
principal and the auxiliary amplitudes. We will show how
a machine learning strategy can be employed based on the
circular causality to simplify coupled cluster 
calculations. In section III we discuss each of the machine
learning models and their performance in terms of accuracy
and time requirements. We will discuss four models, out 
of which, three are primarily regression based, and the
last one is a classification based model. In section IV, 
we conclude our findings with overarching remarks on the
performance of the models.

\section{Nonlinearity in the CC iteration scheme from synergistic perspective, and overview of a new algorithm based on the circular causality relationship:}
\label{last_sem}

There exists a relationship
among the set of most significant and less significant
cluster amplitudes during the nonlinear iteration process.
While the significant amplitudes dictate the iteration 
process in the macroscopic sense, other less significant
amplitudes are simply \textit{enslaved}, and their 
variation gets suppressed. These significant amplitudes,
which are large in magnitude and mostly labelled by the
active orbitals, are interchangeably 
termed as the principal or the driver amplitudes. From the 
viewpoint of the nonlinear discrete time dynamics, they are
the unstable modes and are denoted by $t^L$. The space 
spanned by them are termed as the large amplitude 
subset (LAS), having dimension $n_L$. On the other
hand, the amplitudes with lesser magnitudes than 
a pre-defined threshold have significantly less importance in 
the iteration dynamics, and they are the stable modes of the
time-discrete iteration process. They are termed as the 
auxiliary amplitudes, denoted by $t_S$, which span the 
small amplitude subset (SAS) with dimension $n_S$. Note that 
$n_S >> n_L$. The principal and the auxiliary amplitudes are 
known as master and slave variables in Synergetics.
It was demonstrated that there exists a mapping $F$, 
such that 
\begin{equation}
    F:t^L_k\rightarrow t^S_k
    \label{eq:mapping}
\end{equation}
where $k$ is the discrete time iteration step. We 
established that
one may numerically exploit this mapping to reduce the 
computational time required for the coupled cluster 
calculations. 
\begin{figure}[!h]
    \centering
\includegraphics[width=\linewidth]{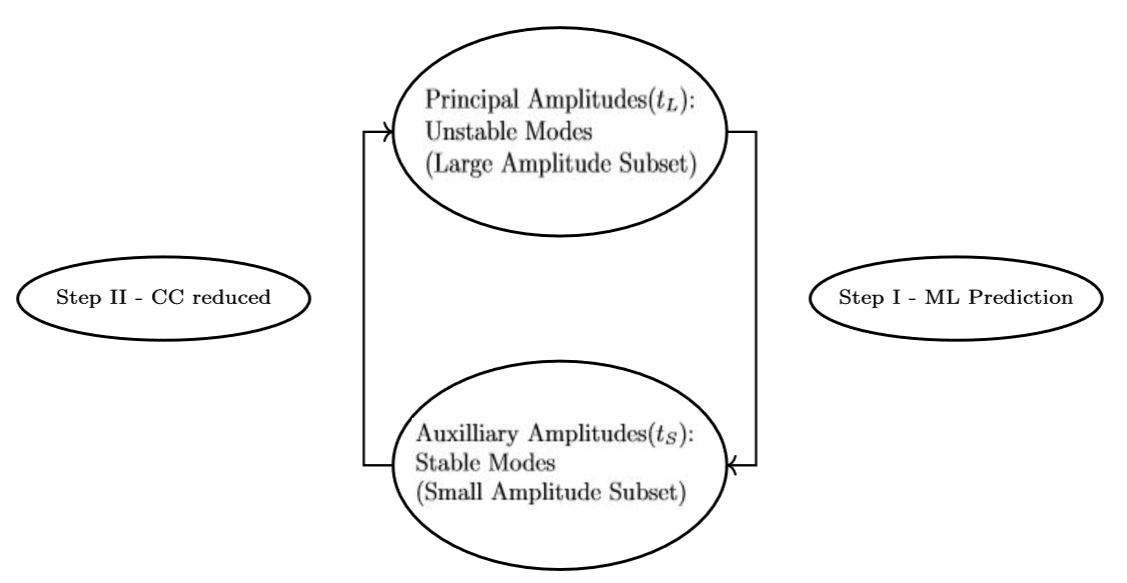}
\caption{General structure of the circular Causality loop: the principal amplitudes are mapped on to the auxiliary amplitudes using a ML model (step-I), and their feedback coupling to obtain the updated set of the driver amplitudes via step-II. This loop is constructed following the initial iteration cycles and training of the relevant ML model.}
    \label{fig:circ-caus}
\end{figure}
In order to establish the mapping $F$, one may modify the
CC iteration scheme based on the circular causality that 
exists between the principal and the auxiliary amplitudes. 
This is shown in Fig. \ref{fig:circ-caus}. However, the 
circular causality presumes that the map $F$ is known. In 
order to establish the mapping, one may use the routine CC
framework to calculate both the sets of amplitudes, 
$t^L$ and $t^S$. After a certain number of \textit{training} iterations,
the principal amplitudes are extracted and they are 
trained to map on to the auxiliary amplitudes for once.
This mapping in our current scheme is done via various
kernelized Regression based ML models where the principal
amplitudes are taken as the independent input variables 
and the auxiliary amplitudes are
the dependent variables that are predicted. For each individual
molecular/atomic calculation, this training needs to be done 
right after the initial iterations only once, and 
there is no requirement of any previously computed data. Our 
model can be trained on the amplitudes on the fly during 
the iteration process. Depending on the desired level of 
accuracy, we will present four different regression models
that have their own pros and cons, and we will discuss 
those in the sections to follow, along with their
exhaustive numerical applications. Once the amplitudes 
are trained and the 
functional form of $F$ which maps $t^L$ on to $t^S$ is 
numerically established, the algorithm runs via the 
circular causality loops as shown in Fig. \ref{fig:circ-caus}.
This consists of two steps: in step-I, all the auxiliary
amplitudes are determined from the principal
amplitudes of the same iteration time step via the ML 
regression model. In step-II, the principal amplitudes 
are updated via the feedback coupling of
both sets of the amplitudes. This circular causality 
loop continues till the principal amplitudes (and hence 
the auxiliary amplitudes, due to the fixed functional
dependence) converge. Note that in step-II, only the 
principal amplitudes are determined through the exact CC
equations, although the auxiliary amplitudes that are plugged 
in are obtained via ML. The computational scaling of this 
step is $n_Ln_o^2n_v2$ as demonstrated by Agarawal 
\textit{et. al.}\cite{agarawal2021accelerating}. 
Since $n_L$
is only a very small fraction even compared to $n_on_v$, 
this step has a significant reduction in scaling from the
usual $n_o^2n_v^4$ of the conventional scheme. Step-I 
involves only a single matrix multiplication, and hence 
it scales negligibly compared to step-II. We will henceforth
refer to this algorithm as the hybrid CC-ML scheme.

The computational time requirement by the overall hybrid 
CC-ML scheme is governed by two major factors: (1) the number 
of conventional iterations $(m)$ used to train the model to 
determine $F$, and (2) the dimension of the LAS, $n_L$.
Simple calculations show that over 90\% of the overall
calculation time is consumed in generation of data 
through the initial iterations and 
training of model. Thus, there is a lot of scope to play
around with different ML models that can predict the 
auxiliary amplitudes with fewer training iterations, and
minimize the required dimension of the the LAS, $n_L$,
without unduly sacrificing the accuracy. 

In a previous publication, Agarawal \textit{et. al.}\cite{agarawal2021accelerating} had studied the
efficacy of KRR based ML model to represent $F$.
This was studied for a few pilot molecules in their 
equilibrium and away from equilibrium geometries, and it
produced highly accurate energies (of the order of 
micro-Hartree ($\mu H$), compared to the canonical CC
calculations). The authors also 
systematically studied the time requirements necessary for 
the overall process, and as expected, it was much less 
than the conventional CCSD scheme. In fact, in order to
obtain a sub-$\mu H$ accuracy in energy, the time 
requirement for the algorithm was similar to lower to 
that of a DIIS
accelerated CC scheme. Thus, the pilot study showed 
extremely promising results, and it required further
benchmarking. In this manuscript, we have thoroughly 
studied the performance of KRR model, both in terms of 
energy accuracy and time requirement, for about 60 
molecular applications with various degrees of electronic
complexity. We would also introduce a few more regression 
models, which would further greatly reduce the requirements 
of the numbers of initial training iterations, and the
LAS dimension. All the models will be thoroughly
benchmarked with numerous numerical applications.

\section{Discussion on various regression based machine learning models and their benchmarking studies}
\label{sec:ML-models}
In this section we will describe various regression based ML
models that have been employed in this work. We have studied
four different regression models to represent the forward
mapping (step-I) of the circular causality loop. 
The models which we would be considering in this 
manuscript are:
\begin{itemize}
    \item Kernel Ridge Regression model (KRR)
    \item Customized Kernel Regression model (CKR)
    \item Polynomial-Kernel Ridge Regression model (PKR)
    \item K-Nearest Neighbors model (KNN)
\end{itemize}
Following the initial iteration cycles, the model is trained 
only once for each molecule, after which the iteration runs 
through the circular causality loop. 
In all the ML models, the input parameters 
are the principal amplitudes, $t^L$, which are taken to be 
the independent variables, and the output is the set of
auxiliary amplitudes $t^S$ which are predicted. While CKR and 
PKR models superficially have the similar working philosophy 
to that of KRR, they differ from KRR in the detailed
structure of the Kernel matrix. Hence, the underlying 
mathematical
structure of all these three models will be discussed together 
in the section that deals with KRR. In the subsequent sections 
dealing with CKR and PKR models, we will only present the
Kernelization techniques which distinguish them from KRR. 
In the following, we will briefly 
present the working principles of all the different models 
and would also present the benchmarking applications along 
with, in comparison to the canonical CCSD results. For each
molecule, both the canonical CCSD and hybrid CC-ML calculations
were done in the same machine for proper comparison of the 
relative computation timing.

\subsection{The Kernel Ridge Regression Model:}
\label{ssec:ML-models-KRR}

\subsubsection{Formal aspects and structure of the working equations:}

In this subsection, we will discuss the main ideas behind the
Kernel Ridge Regression model. As mentioned before, this model
also forms the base for two other models under consideration,
namely the Customized Kernel Regression and the Polynomial
Kernel Ridge Regression. 
The KRR model is strictly based on the linear regression
algorithm. A linear regressor fits the dependent auxiliary
amplitudes as a linear function of the independent principal
amplitudes. The training of model essentially translates to
finding appropriate coefficients in a linear function of
principal amplitudes to give auxiliary amplitudes. As one
would expect, the accuracy of the function would be directly
dependent on the number of training iterations $m$.

As mentioned previously, during the training cycles, the CC
equations are iterated in the full space spanned by
$\{t^L\oplus t^S\}$  amplitudes and these exact amplitudes
are used to train the model. During the training, the entire space of $\{t^L\oplus t^S\}$ is processed into two matrices of $T^S$ and $T^L$, where $T^L$ is defined as 
\begin{equation}
T^L = 
  \begin{pmatrix}
    1 & t^L_{11} & t^L_{21} & ... & t^L_{{n_L} 1}\\
    1 & t^L_{12} & t^L_{22} & ... & t^L_{{n_L} 2}\\
    1 & t^L_{13} & t^L_{23} & ... & t^L_{{n_L} 3}\\
    ... & ... & ... & ... & ...\\
    1 & t^L_{1m} & t^L_{2m} & ... & t^L_{{n_L} m}\\
\end{pmatrix}  
\label{eq:krr_tl_input}
\end{equation}.
Here each row signifies the $n_L$ independent principal 
cluster amplitudes for a given iteration. With $m$ number of
training iterations performed to construct the $T^L$ matrix,
it is of the dimension $m\times (n_L+1)$. The extra column in 
$T^L$ is added to take care of the intercept term. 
A generalized linear fit can be written in the form of 
$T^S = T^L \beta$, where $\beta$ is the coefficient matrix 
of the linear functions of $t^L$. It can be trivially 
defined as: 
\begin{equation}
\beta = 
  \begin{pmatrix}
    \beta_{01} & \beta_{02} & \beta_{03} & ... & \beta_{0 {n_s}}\\
    \beta_{11} & \beta_{12} & \beta_{13} & ... & \beta_{1 {n_s}}\\
    \beta_{21} & \beta_{22} & \beta_{23} & ... & \beta_{2 {n_s}}\\
    ... & ... & ... & ... & ...\\
    \beta_{{n_L} 1} & \beta_{{n_L} 2} & \beta_{{n_L} 3} & ... & \beta_{{n_L} {n_S}}\\ 
\end{pmatrix} 
\label{eq:krr_beta}
\end{equation}
where $n_S$, as mentioned before, is the number of the
elements in the dependent SAS amplitudes for a given
iteration. Thus the matrix $T^S$ is of the dimension 
$m\times n_S$, defined as:
\begin{equation}T^S = 
     \begin{pmatrix}
    t^S_{11} & t^S_{21} & ... & t^S_{n_S 1}\\
    t^S_{12} & t^S_{22} & ... & t^S_{n_S 2}\\
    t^S_{13} & t^S_{23} & ... & t^S_{n_S 3}\\
    ... & ... & ... & ...\\
     t^S_{1m} & t^S_{2m} & ... & t^S_{n_S m}\\
\end{pmatrix}  
\label{eq:krr_ss_input}
\end{equation}
Starting from a guess coefficient matrix, $\hat{\beta}$,
the optimized coefficients may be obtained by minimizing 
the loss function $\eta^T \eta$. Here $\eta$ 
is the error function defined by:
$\eta = T^S - \hat{T^S}$, where 
$\hat{T^S}=T^L\cdot \hat{\beta}$ is the predicted auxiliary
amplitude matrix obtained during the training cycles. With 
$m$ training iterations, $\hat{\beta}$ can be optimized 
to give $\beta$, which will be used for prediction in the
subsequent iterations.

One may further increase the power of the independent 
principal amplitudes to quadratic, cubic etc. to  
capture the nonlinear features of the trajectory. This 
leads to increase the dimension on the independent 
variable space, and hence there's a better chance of getting a
better fit. One thus increases the independent variable space 
from a dimension of $n_L$ to a higher dimensional space by 
including polynomial forms of $t^L$ and treat each terms as
independent variable. Let the higher dimension space with non linear terms be connected to the feature vectors by a map $\phi$, which can be defined as: 
\begin{equation}
    \phi: \{t^L_\mu, \mu \in n_l\} \rightarrow\{t^L_\mu,t^L_\mu t^L_\nu,...,(t^L_\mu)^d, \mu,\nu\in n_l\}
    \label{eq:lin_to_non_lin_kernel}
\end{equation}
where $d$ is the degree of the polynomial. The loss function now becomes $\hat{\beta_\mu}=\phi^T (\phi \phi^T)^{-1} (T^S_\mu)$.
The whole operation, if done directly, is expensive and one 
usually makes use of the Kernelization technique instead, which 
allows evaluation the expression without explicit 
knowledge of the function $\phi$.

According to the Mercer theorem\cite{minh2006mercer}, one may
define the Kernel Function $K=\phi \phi^T$
for every symmetric positive definite matrix.
Thus, the loss function now becomes  $\hat{\beta_\mu}=\phi^T (K)^{-1} (T^S_\mu)$. 
After sufficient training of the model, the auxiliary SAS 
amplitudes for the $i$-th iteration ($i>m$) ($t_i^S$) are
predicted using only the optimised coefficient matrix and 
the LAS amplitude vector of the same iteration ($t_i^L$).
\begin{equation}
    t_{\mu,i}^S (predicted) = \phi(\{t^L_i\}) \hat{\beta}_\mu
\end{equation}Note that the effect of $t^S $ or previous iterations is included in the $\hat{\beta}$ matrix. 

To bypass the usage of $\phi$, a new kernel function 
$K_L$ is defined, which takes all the previous training amplitudes, 
and the $t_i^L$ to predict the $t^S_i$. Thus, 
$K_L = \phi. (\phi(t^L_i))^T$, 
where the right most $\phi$ appearing in 
the above equation is a 
function of $t_i^L$.
\begin{equation}
    t^S_{pred} = (K_L)^TK^{-1}T^S
    \label{eq:KernelReg}
\end{equation}
Here the subscript indicates that these are the 
predicted SAS amplitudes. 

To avoid unphysical underfitting or overfitting due to 
erroneous weight in the training data set, 
a regularization parameter is introduced 
which penalizes the model each time a certain
term gets unphysical weight. 
Thus one may modify Eq \ref{eq:KernelReg} by 
adding a regularization term as \begin{equation}
     t^S_{pred} = (K_L)^T(K+\lambda I)^{-1}T^S
     \label{eq:ridge}
\end{equation}
Following the convention, we denote the 
regularization parameter with $\alpha$. Here
$\lambda=\alpha/2$. The value of 
$\lambda(\alpha)$ manages overfitting at the
cost of rate of learning. A small value
of $\alpha$ trains a model very quickly and often 
overfits the data points, which could result in 
slight inaccuracies with few training data 
sets. On the other hand, a large $\alpha$ 
slows down the learning process, and the 
model takes a larger number of training data 
sets to produce accurate results. Thus the value of $\alpha$ requires tuning. 

In Eq. \ref{eq:ridge}, we note that the 
quantity $(K+I \lambda)^{-1}T^S$ can be computed 
only once for all after the training data set
is produced. Thus all the $t^S_i$ amplitudes 
may be computed each iteration via a
single matrix multiplication of $(K_L)^T$ and $(K+I \lambda)^{-1}T^S$ instead of algebraically or diagrammatically
solving the coupled cluster equations, which results in
savings in the overall computational time. We must also 
note that all these processes are already standardized 
in scikit \cite{scikit-learn}, and one has to only 
supply $T^L$ and $T^S$ for training.

In an earlier paper, we had presented a few pilot molecular 
applications to show the potential of the hybrid CC-ML(KRR) 
method. However, a more thorough analysis of the model on a 
large number of molecules with varying electronic complexity is
necessary to show the efficacy of the method with statistical
significance. We have employed the KRR model to study the
ground state energetics of about 60 cases with
different molecules/geometries/basis sets. In the next 
subsection, we will study the efficacy of the KRR model in
terms of accuracy in predicted energy compared to CCSD, 
and the associated time requirements.

\subsubsection{Assessment of the performance of KRR model:}
\begin{figure*}
    \centering
    \includegraphics[width=\textwidth]{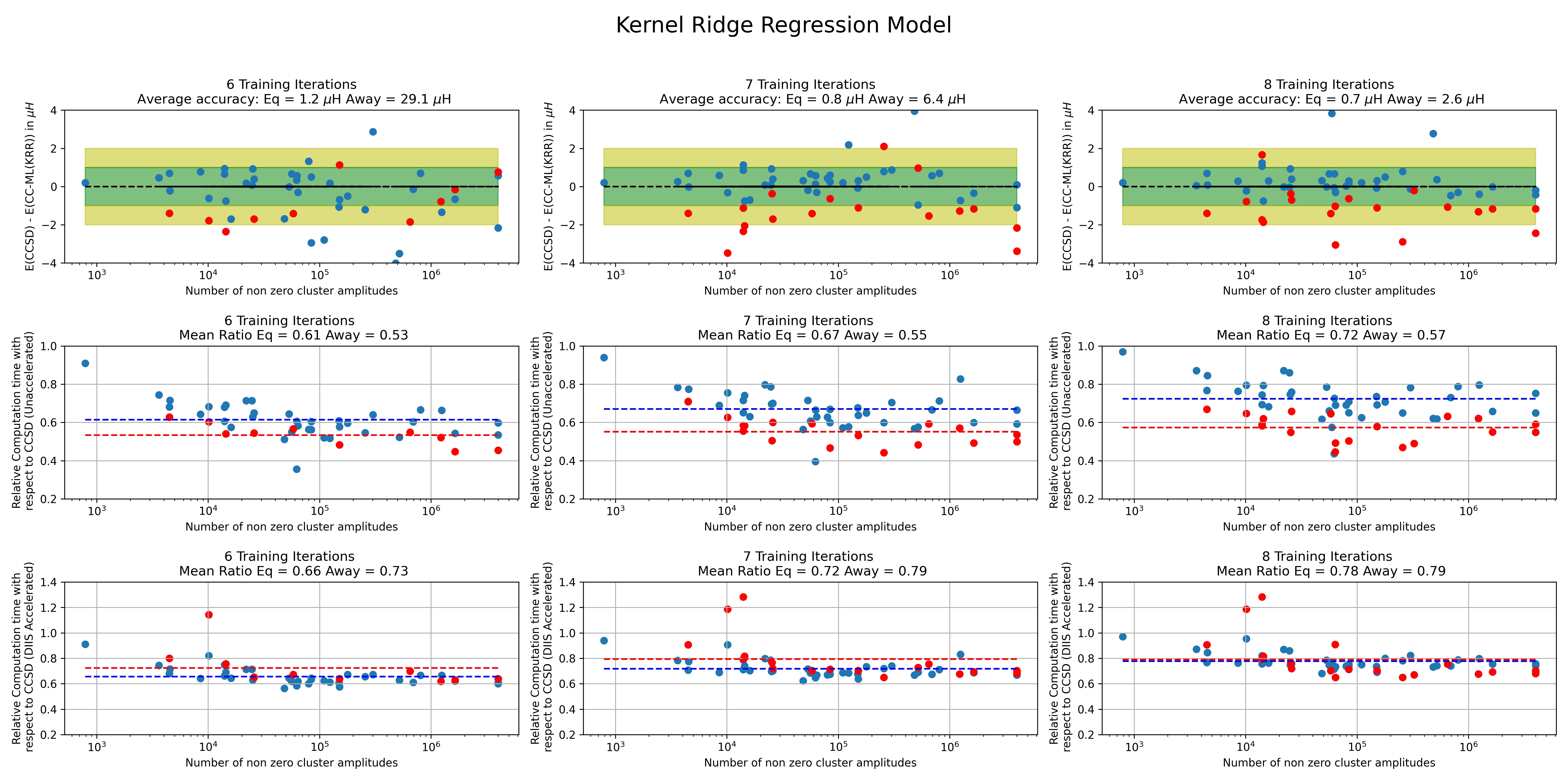}
    \caption{Performance of hybrid CC-ML(KRR) model: the x-axis on all sub-plots are arranged in the increasing number of the non-zero cluster amplitudes, which is taken as a measure of the system size. The blue dots in all sub-plots are for the molecules in their respective equilibrium geometries, and the red dots are for the molecules in stretched geometries. The first row shows the difference in energy between the CC-ML(KRR) and exact CCSD. The black horizontal line is the reference, and the green horizontal bar signifies the convergence accuracy of $\pm$1$\mu$H. The second row shows the relative time taken by CC-ML(KRR) with respect to the exact CCSD (without DIIS acceleration). The red and blue lines show the mean fraction of the computational timing. The third row shows the time taken by CC-ML(KRR) with respect to DIIS accelerated CCSD scheme. \enquote{Eq} and \enquote{Away} in the sub-figure captions refer to the molecules in their equilibrium and away from equilibrium geometries, respectively. The three columns denote the number of training iterations employed.}
    \label{fig:krr-bench}
\end{figure*}
The accuracy of any regression method largely depends on the 
number of iterations used to train the model. As expected,
larger the training iterations employed, better would be the 
accuracy. In the first row of Fig \ref{fig:krr-bench}, we have
plotted the difference in the energy obtained via canonical
CCSD and our hybrid CC-ML(KRR) model for 6, 7, and 8 training 
iterations for 60 different combinations of molecules, 
basis sets and geometries. With 6 training
iterations, for most of the molecules in their respective
equilibrium geometries, the hybrid CC-ML(KRR) predicts energy
which is accurate up to 2 $\mu H$ to the exact CCSD energy.
However, there are quite a few cases, particularly for 
molecules in distorted geometries, for which 6 training
iterations are not sufficient to accurately determine $F$, 
and they provide results which are off by a large margin. 
They are not included in the scale of the plot. However, 
with higher number of training iterations, the difference
between the 
energy obtained via canonical CCSD and hybrid CC-ML(KRR) 
tend to converge more towards the middle. This is particularly 
true for molecules in away from their equilibrium geometries,
which are shown by the red dots. The mean absolute 
deviation (MAD) for molecules in away from equilibrium regions
converge very fast with the number of training iterations 
with MAD=29.1 $\mu H$ with $m=6$ to MAD=6.4 $\mu H$ with 
$m=7$ to MAD=2.6 $\mu H$ with $m=8$. For the molecules in their
equilibrium geometries, even with 6 training iterations, 
MAD of 1.2 $\mu H$ is observed, which further systematically
improves as we increase the number of training iterations.
Overall, irrespective of the molecular geometry, with $m=7$,
the CC-ML(KRR) predicts energy within $\pm 2$ $\mu H$, which
are shown by the yellow band.

In the second row of Fig \ref{fig:krr-bench}, we have plotted 
the fraction of the time taken by the hybrid CCSD-ML(KRR)
scheme compared to the conventional CCSD calculations. 
With $m=6$, the hybrid 
CC-ML(KRR) scheme takes at around 61\% time compared to the
conventional CCSD calculations for molecules in the equilibrium
geometries. For the molecules with away from equilibrium 
geometries, the hybrid CC-ML(KRR) on an average takes around
53\% of the total computational time to that required for the 
an CCSD. As one includes more training iteration, the time
requirement increases, and the scattered points shift upwards.
One may also note that the 
blue points (which are assigned for molecules with equilibrium 
geometries) appear higher than the red points (for molecules 
with stretched geometries). This is due to the fact that the
molecules in equilibrium geometries take much lesser number of 
CCSD iterations to converge, and hence the training cycles 
take a large fraction of the overall calculation. In 
stretched geometries, the CCSD calculations take longer time 
to converge, which means that for the hybrid CC-ML scheme, 
the cost of initial iterations and training of the model 
is small relative to the total computation time, and hence 
the red dots appear consistently below the 
blue points. Note that with 8 training iterations, the computation time ratio is only
0.72 and 0.57 for equilibrium and away from equilibrium  geometries respectively, relative to the canonical CCSD
calculations. In the case where the CCSD scheme is accelerated
via DIIS, the ratio is slightly higher, as expected and 
observed from the third row of Fig \ref{fig:krr-bench}. 
However, the hybrid CC-ML(KRR) still outperforms DIIS
accelerated CCSD in terms of the required computational time
for most case irrespective of the molecular geometry. 

\subsection{Customized Kernel Regression Model:}
\label{ssec:crm}

While the hybrid CC-ML(KRR) model is very stable and accurate, 
it requires a larger number of training set. Moreover, one
needs to start with an unknown polynomial form of the 
coupling map. Moreover, in the CC-ML(KRR) model, even starting
from the first order guess amplitudes, one needs to discard 
a couple of initial iterations as the iteration pattern does 
not get stabilised to be accurately predicted by the model.
This makes the KRR model quite data intensive and takes high
number of initial iterations to train the data for sufficiently
accurate result. Moving towards a map that is more consistent
with the numerical pattern of the CC iteration scheme is 
expected to reduce the number of training iterations and the 
required dimension of the independent amplitude space. We 
have generated a new kernelization technique that is more
consistent with the CC iteration scheme. As regularization 
often slows down the learning process, one may remove the 
Ridge regularization from the Eq \ref{eq:ridge}.
In order to speed up the calculations and get as much 
features as possible from very less amount of input
parameters, we go back to Eq. \ref{eq:KernelReg}, and define 
our own custom made Kernel function. Note that KRR uses a 
ready-made polynomial kernel available in the standard 
library; however, CKR is based on our own kernel matrix, 
which we integrated with the library of the Scikit.

Following a similar philosophy as KRR, we define the input 
training matrix $T^L$ and $T^S$ exactly the same way as 
defined in Eqs \ref{eq:krr_tl_input} and \ref{eq:krr_ss_input}.
To extract the input features from the $T^L$ and $T^S$ 
matrices, we define two different feature matrices
\begin{itemize}
    \item The Squared Root Euclidean Distance Matrix (EDM), 
    or what has simply been
    defined as distance matrix $D$ in our earlier works 
    \cite{agarawal2020stability,agarawal2021accelerating}. The 
    elements of the distance matrix $D$ is defined as 
    \begin{equation}
        d_{ij}=||\Vec{x_i}-\Vec{x_j}||,
        \label{eq:sqrt_eucl}
    \end{equation}
    where $\Vec{x_i}$ is defined as the $i-th$ row of $T^L$.
    
    \item Linear Kernel ($Q$, elements denoted as $q_{ij}$) is defined as 
    \begin{equation}
        Q=T^L(T^L)^T
        \label{eq_Q_old}
    \end{equation}
\end{itemize}
Both of the matrices are symmetric and of the size $m \times m$,
where $m$ is, as usual, the number of the training data sets. The 
EDM and the linear kernel matrix are the input to the kernel,
and they work as feature extractors from the training data. 

With these feature matrices, a new kernel is constructed with
the combination of an exponential and sinusoidal parts. The 
the kernel $K$ with elements $k_{ij}$, $i,j==\{1,2,...p\}$ 
can be written as 
\begin{equation}
    k_{ij}=exp(-sin(g_{ij}+l_{ij}))
    \label{eq:crm_kernel}
\end{equation}
where,
\begin{equation}
    g_{ij}=exp(-\gamma*d_{ij}), l_{ij}=log(q_{ij})
    \label{eq:crm_components}
\end{equation}
where $\gamma$ is a hyper-parameter, which can be optimized. 

Once the model is trained through the initial cycles, 
we first calculate the new kernel matrix, $K_L$ with the new 
input values of the principal amplitudes through the 
new vector $T^L_{new}$. 
In order to do so, we first calculate the feature extraction
matrices, where the new squared root EDM $D_L$ is defined as: 
\begin{equation}
    D_{L\textunderscore i} = \sqrt{\sum_{k=1}^n(t^L_{(new)ik}-t^L_{ik})^2}\;,i=\{1,2,...,m\}
    \label{eq_D_new}
\end{equation}
and 
the linear kernel matrix $Q_L$ as:
\begin{equation}
    Q_L=T^L (T^L_{new})^T
    \label{eq_Q_new}
\end{equation}
In Eqs. \ref{eq:sqrt_eucl} and \ref{eq_Q_old}, we now replace
the elements of $D$ and $Q$ matrices by the elements of the 
new feature matrices $D_L$ and $Q_L$ from Eqs. \ref{eq_D_new} 
and \ref{eq_Q_new}. Finally, using Eq. \ref{eq:crm_kernel},
we get the predicted auxiliary amplitudes as:  
\begin{equation}
    t^S_{pred} = (K_L)^TK^{-1}T^S
    \label{eq:crm_pred}
\end{equation}
Although the hybrid CC-ML(CKR) model performs quite well 
in most of the cases, it is observed that due to the
sinusoidal nature of the kernel matrix, sometimes for some 
highly fluctuating larger amplitudes, combined with 
a bad choice of $\gamma$, may make ($g_{ij}+l_{ij}$)
inside the exponential lose its monotonic nature.
In that case, $K$ may contain very small diagonal 
value, and taking its inverse via Eq. 
\ref{eq:crm_pred} makes the whole model to explode 
with a very large value of the predicted auxiliary 
amplitudes. Forceful regularization procedure on the
diagonal values does not work well, as it changes the
overall nature of the kernel matrix, which later 
gives poor accuracy. One of possible way is to remove
more fluctuating larger amplitudes from the training 
matrix itself before training. However, the 
consequence of the removal of the highly fluctuating 
large amplitudes results in poorer accuracy of the
overall model. Furthermore, extracting the highly
oscillating large amplitudes beforehand is a highly
nontrivial exercise in ML. A more stable model in 
this regard will be a subject of our forthcoming
publication. In our model, to prevent this from
happening, we integrated a separate algorithm which
keeps track of the $K_L$ matrix, and prevents it from
changing any of its element value by more than 
‘0.02’ by pushing it backwards. This whole process
makes the learning process slow and does not give
the desired accuracy. Nonetheless, this procedure 
prevents the whole code from catastrophic breakdown.
With this caveat, the results for the CKR model 
are discussed below. 

\subsubsection{Assessment of the performance of CKR model:}
\begin{figure*}
    \centering
    \includegraphics[width=\textwidth]{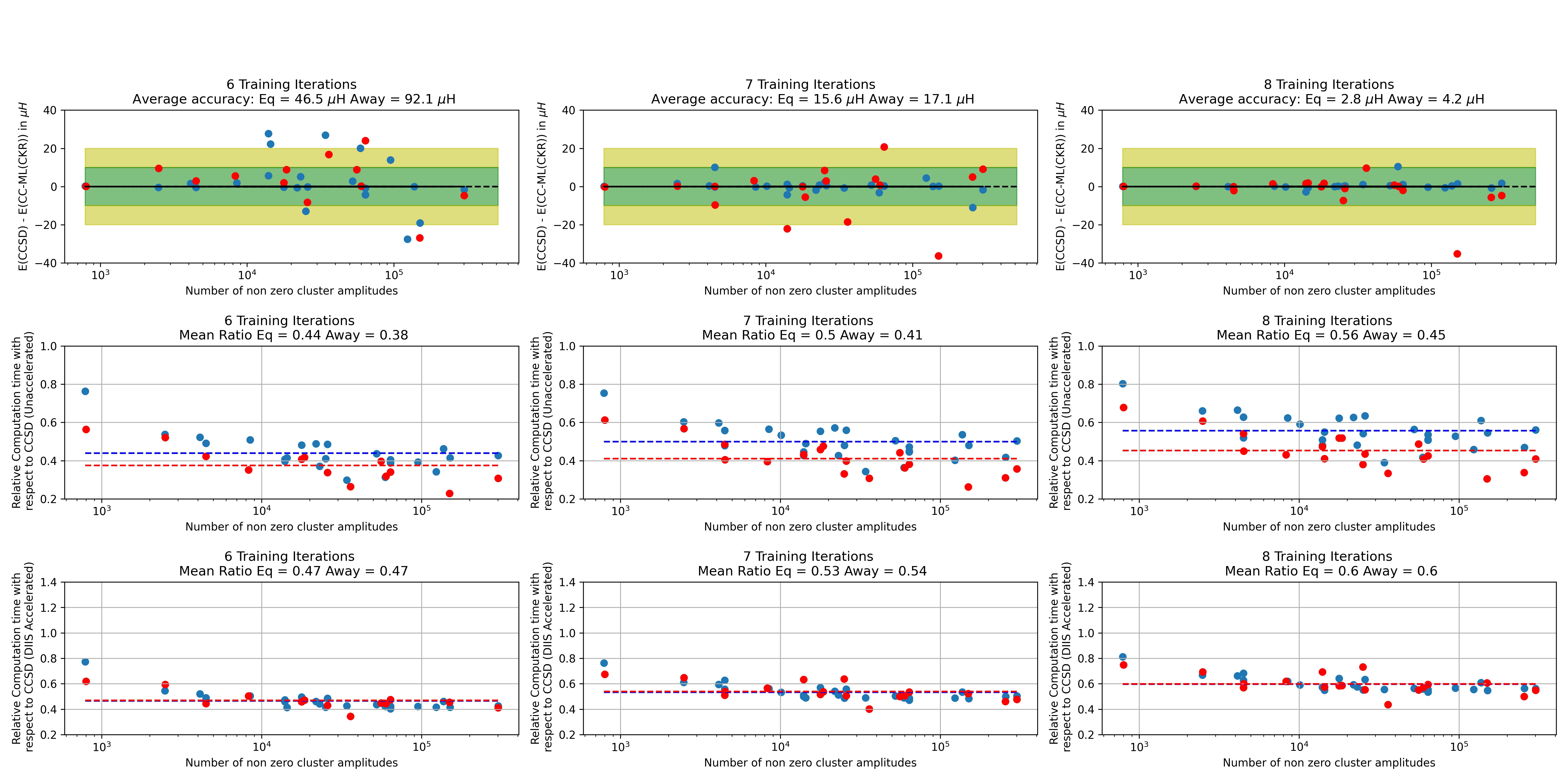}
    \caption{Performance of CC-ML(CKR). The first row shows the difference in energy between the CC-ML(CKR) and the exact CCSD with the green bar showing an energy accuracy of $\pm 10 \mu H$. The second and third rows show the relative computational time of CC-ML(CKR) with respect to CCSD (without DIIS acceleration) and DIIS accelerated CCSD schemes, respectively. The three columns denote the number of training iterations employed for training. All other details are same as Fig.\ref{fig:krr-bench}.}
    \label{fig:crm_bench}
\end{figure*}

It is evident from the previous discussion that the CKR 
method is a high-risk/high-profit model where for 
almost all the cases under consideration, one gets highly
accurate result with less computation time compared to KRR.
Like all other models considered in this manuscript, the 
accuracy largely depends on the number of training iterations.
However, unlike KRR, the customized kernelization predicts 
the trajectory with less number of training iterations and
discarding the couple of initial iterations is not needed.
This saves a significant computation time due to less training
iterations. Moreover, The CKR model requires a much smaller 
size of the independent variables compared to KRR,
($n_{L_{KRR}}>>n_{L_{CKR}}$). This means the major
computational bottleneck of $n_Ln_o^2n_v^2$ scaling in step-II
is significantly reduced. This thus allows a faster loop over  
the LAS elements, reducing the overall computation time
significantly over KRR. Note that with a higher value of $n_L$,
although the computation time increases slightly due to the
$n_Ln_o^2n_v^2$ scaling in step-II, the results become 
significantly more accurate. 

The performance of hybrid CC-ML(CKR) model is shown in 
Fig. \ref{fig:crm_bench} for various numbers of the training 
iterations for about 52 different molecules/basis/geometries.
With $m=6$, the model is significantly less accurate than 
KRR with MAD of 46.5 $\mu H$ and 92.1 $\mu H$ for molecules 
in equilibrium and away from equilibrium geometries 
respectively, from the 
canonical CCSD calculations. However, the model has a sharper 
convergence with respect to the number of the training cycles 
compared to KRR. With $m=8$, the MAD observed with the CKR
model is only 2.8 $\mu H$ for the molecules in equilibrium 
geometry, while for the molecules with away from equilibrium
geometry, the MAD is about 4.2 $\mu H$. As we mentioned before,
unlike KRR, one does not need to discard the initial iterations
to train the model, and hence the training can be performed with
less number of iterations effectively. This amounts to 
significant reduction in the computation time than KRR. This
is reflected in the second and third row panels of 
Fig. \ref{fig:crm_bench} where we have plotted the relative 
time requirement compared to the conventional CCSD. Note that
for $m=8$, the average time requirement to obtain energy of
$\mu H$ precision is only 56\% of that of the canonical CCSD
calculation for molecules in equilibrium. To achieve a similar
accuracy for the molecules with distorted geometry, the
average time requirement is about 45\% compared to canonical
CCSD. If the canonical CCSD calculations are accelerated by
DIIS as is done in most of the cases, the time requirement 
for the hybrid CC-ML(CKR) slightly goes up to 60\% on an 
average to the DIIS accelerated CCSD calculation. However,
one should monitor the predicted energy from CKR as in few
rare cases, the model may get unstable as mentioned before
producing somewhat inaccurate energy.

\subsection{Polynomial Kernel Ridge Regression Model:}
    
The Polynomial Kernel Ridge Regression model has a similar
structure to that of the CKR model presented in the previous 
section, and hence again, we only present the essential 
aspects which distinguish PKR from KRR. While the KRR model 
is linear by default, there is supposed to have more
variational flexibility by introducing the
nonlinear terms in the kernel function. To proceed along this 
direction, we define the input $T^L$ and $T^S$ matrices the 
same way as Eq. \ref{eq:krr_tl_input} and Eq.
\ref{eq:krr_ss_input}. Furthermore, we define the linear 
kernel, $Q$ as Eq. \ref{eq_Q_old}. We then introduce our
polynomial kernel $K$ as 
\begin{equation}
    K=(\gamma Q +C)^o
\label{eq:kernel-poly}
\end{equation}
where $C=1$ is a constant, and $o=3$ is the order of the 
polynomial. These values of $C$ and $o$ are taken by default 
in the ML library\cite{scikit-learn}. Like the previous two
models, this kernel is now used in training of the amplitudes.
The prediction is done by a similar technique used in
the KRR model using Eq. \ref{eq:ridge}. The most
fundamental difference between the KRR and PKR is in
the choice of the kernel function: while the former
uses a linear regression type fitting, the latter
uses polynomial function, in this case of degree 3,
for a better accuracy.

\subsubsection{Assessment of the performance of PKR model:}
\begin{figure}[!b]
    \centering
    \includegraphics[width=\textwidth]{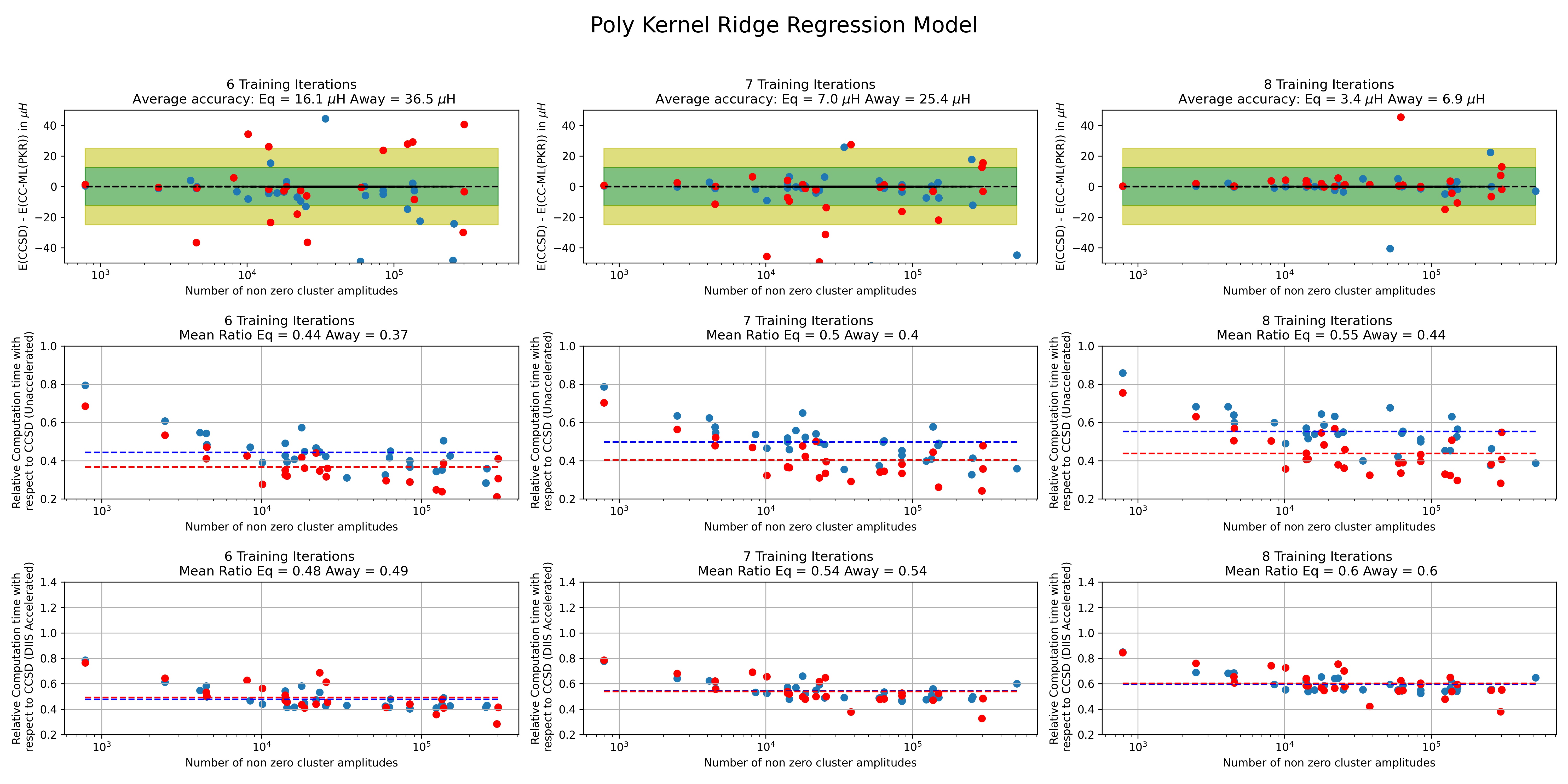}
    \caption{Performance of CC-ML(PKR). The first row shows the difference in energy between the CC-ML(PKR) and the exact CCSD with the green bar showing an energy accuracy of $\pm 12.5 \mu H$. The second and third rows show the relative computational time of the model with respect to un-accelerated and DIIS accelerated CCSD schemes, respectively. The three columns are for three different training iterations ($m$=6, 7, 8). All other details are same as Fig. \ref{fig:krr-bench}.}
    \label{fig:poly-kr}
\end{figure}

The PKR method is a variant of the parent KRR model.
This model, as mentioned previously, increases the
dimension of the feature matrix, which allows us a
better fit of the auxiliary amplitudes in terms of 
the principal amplitudes. PKR model, as a result, is
highly accurate. Howeverm, like CKR, this too does
not discard any initial iterations, and hence has a
similar requirement of the number of training 
cycles as that of CKR. Since the number of 
independent degrees of freedom increases due to the
polynomial structure of the kernel via Eq. 
\ref{eq:lin_to_non_lin_kernel}, a small 
dimension of the LAS, $n_L$, is sufficient to achieve
high accuracy. The performance of the hybrid 
CC-ML(PKR) model is shown in Fig. \ref{fig:poly-kr}
as a function of different number of the training
iterations for about 60 different molecules/basis/geometries.
With $m=6$, the model is less accurate than 
KRR with MAD of 16.1 $\mu H$ and 36.5 $\mu H$ for
molecules in equilibrium and away from equilibrium 
geometries respectively, from the canonical CCSD
calculations. This is substantially better than CKR
model with equal number of training iterations. Even
though the model has a sharper convergence of the 
estimated energy than KRR, it is not as sharp as CKR.
This is presumably due to the fact that the PKR 
model, by construction, starts with an unknown
polynomial mapping, and also it has a ridge
regularization term. With $m=8$, the MAD for the PKR
model goes down to 3.4 $\mu H$ for the molecules in 
equilibrium geometry, and 6.9$\mu H$ for those in the
away from equilibrium geometry. PKR, due to its 
inherent polynomial structure, does not require to 
discard any initial iterations to construct the 
kernel, and hence there is a significant reduction 
in the computation time than KRR. Moreover, PKR 
requires a very small dimension of the principal 
amplitude space, $n_L$, like CKR. This leads to
the time requirement of PKR very similar to CKR.
Following discussions from previous models, we note 
that for $m=8$, the average time requirement is only
55\% of that of the canonical CCSD
calculation for molecules in equilibrium. To achieve 
a similar accuracy for the molecules with distorted
geometry, the average time requirement is about 
44\% compared to canonical CCSD. If the canonical 
CCSD calculations are accelerated by
DIIS, the time requirement for the hybrid CC-ML(PKR)
slightly goes up to 60\% on an average to the DIIS 
accelerated CCSD calculation with $m=8$. 
The model is, however, 
much more stable than CKR due to inclusion of 
regularization terms via Eq. \ref{eq:ridge}. 

\subsection{K-Nearest Neighbours Regression Model:}

The K-nearest neighbor model, strictly speaking, is a
classification based method and not a regression method.
However, it can be modified slightly to turn it into an
approximate regression model, as we briefly discuss below.
Traditionally, the KNN is used in classify a quantity 
into groups looking at nearest points through features
such as Euclidean or Manhattan Distance. Here, we have 
used the Euclidean Distance as the measure of the 
closeness. The neighbours are defined as the individual
training iterations. Therefore, with $m$ training
iterations, we get $m$ neighbours, each neighbour 
having an input principal amplitude feature matrix defined as 
$ T^L_j=\{t^L_{ij}, i\in (1,n_L)\}$, and output as
the remaining auxiliary amplitudes 
$ T^S_j=\{t^S_{ij}, i\in (1,n_S)\}$, 
where $j$ is the number of the training iteration. 
At the end of the training iterations, we have 
$m$ feature vectors and $m$ output
vectors. After starting the reduced iteration, we
calculate the $t^L_{new}$ using CC via step-II of the circular 
causality loop, and the $t^S_{new}$ is predicted as a 
average of the $k$ nearest neighbors, and it's equation 
is given as 
\begin{equation}
    t^S_{new}=\frac{1}{k}\sum_i^k T_i^S
\end{equation} 
where the nearness is defined as the Euclidean distance 
between $t^L_{new}$ and $t^L$ from earlier iterations.
As is clear, the 'nearness' of a point is completely
arbitrary and, we noticed that at just one neighbor,
the accuracy nearly saturated and did not increase with
the increase in the number of neighbors. It is not 
always necessary that the accuracy would increase with
increase in neighbors as it may lead to overfitting. 
With these ideas in mind, and through a well established
ML library\cite{scikit-learn} we converted our ML model 
to be of KNN-Regression form. In the subsequent paragraph,
we demonstrate the performance of the model over several
test cases.

\subsubsection{Assessment of the performance of KNN model:}
\begin{figure}[!t]
    \centering
    \includegraphics[width=\textwidth]{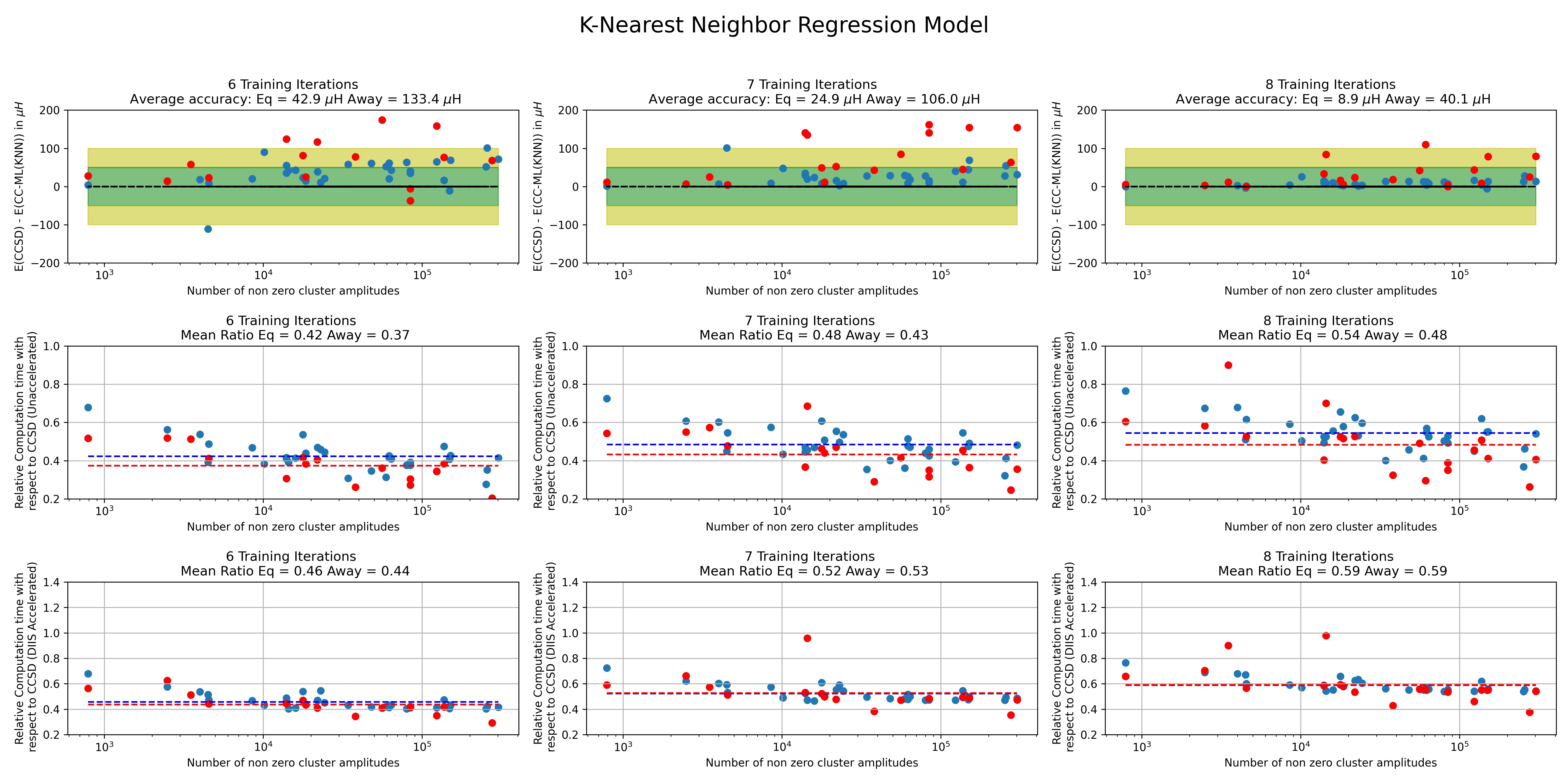}
    \caption{Performance of CC-ML(KNN): the first row shows the difference in energy between the CC-ML(KNN) and exact CCSD with the green bar showing an energy accuracy of $\pm 50 \mu H$. The second and third rows show the relative computational time of the model with respect to un-accelerated and DIIS accelerated CCSD schemes, respectively. The three columns are for three different training iterations ($m$=6, 7, 8). All other details are same as Fig. \ref{fig:krr-bench}.}
    \label{fig:knn-bench}
\end{figure}
In Fig \ref{fig:knn-bench}, we have presented the
performance of KNN for several molecular systems in their
equilibrium and away from equilibrium geometries. Clearly,
the model doesn't perform nearly well as compared
to the other robust regression models discussed previously. While
the average deviation improves sharply by taking 
high number of training iterations, with fewer training 
iterations, the model performs poorly. This contradicts
the basic idea of supervised machine learning where we
target to model the CC iteration convergence with only a
few number of training cycles. This is primarily due to 
it being a classification model, rather than a regression
based model. While the process of training remaining 
the same, which guarantees that it takes approximately 
similar computational time to CKR and PKR, the accuracy 
is not nearly as good as the previously discussed models. 
We must also note that being a rudimentary model, it 
provides us with a base case, and any sophisticated 
machine learning model would be an improvement over it.
Thus the use of such a classification based model is not
recommended to simulate the CC iteration series.

\section{Discussion and summary:}
\begin{table*}
\centering
\begin{tabular}{|p{3.5cm}|M{1cm}|M{1cm}|M{1cm}|M{1cm}|M{1cm}|M{1cm}|M{1cm}|M{1cm}|M{1cm}|}
\hline 
\multirow{3}{*}{Parameters}&\multirow{3}{*}{$n_{train}$}& \multicolumn{8}{C{9.6cm}|}{Machine Learning Models}\\ \cline{3-10}
&& \multicolumn{2}{C{2.4cm}|}{KRR} &\multicolumn{2}{C{2.4cm}|}{CRM} & \multicolumn{2}{C{2.4cm}|}{Poly-KR}& \multicolumn{2}{C{2.4cm}|}{KNN}\\ \cline{3-10}
& &  Eq. & Away& Eq. & Away& Eq. & Away& Eq. & Away \\ \hline
\multirow{3}{*}{Accuracy \newline ($\mu H$)}& 6 &1.2 &29.1 &46.5 & 92.1&16.1 &36.5&42.9 &133.4 \\ \cline{2-10}
& 7 &0.8 &6.4 & 15.6&17.1 & 7.0& 25.4& 24.9& 106.0\\ \cline{2-10}
& 8 &0.7 &2.6 &2.8 &4.2 & 3.4&6.9 & 8.9&40.1 \\ \hline
\multirow{3}{*}{Time/CCSD(UA)}& 6 &0.61 &0.53 &0.44 &0.38 &0.44 &0.37 &0.42 &0.37 \\ \cline{2-10}
& 7 & 0.67&0.55 &0.5 &0.41 & 0.5& 0.4&0.48 &0.43 \\ \cline{2-10}
& 8 & 0.72 & 0.57& 0.56& 0.45& 0.55& 0.44&0.54 &0.48 \\ \hline
\multirow{3}{*}{Time/CCSD(DIIS)}& 6 &0.66 &0.73 &0.47 &0.47 &0.48 &0.49 &0.46 & 0.44\\ \cline{2-10}
& 7 & 0.72& 0.79& 0.53& 0.54& 0.54& 0.54&0.52 & 0.53\\ \cline{2-10}
& 8 & 0.78& 0.79& 0.6&0.6 &0.6 &0.6 &0.59 & 0.59\\ \hline

\end{tabular} 
\caption{Performance of the different Machine Learning models. The accuracy is defined as the absolute difference of the energy obtained by the hybrid CCSD-ML model and the canonical CCSD. The Time/CCSD(UA) denotes the fraction of time taken by the CC-ML model with respect to the unaccelerated CCSD scheme. The Time/CCSD(DIIS) denotes the fraction of time taken by the CC-ML model with respect to an DIIS based CCSD method. All the models, particularly those based on regression, provide excellent accuracy with substantial savings in computational time.}
\label{tab:results}
\end{table*}
Through this work, we have thoroughly benchmarked some of 
the common supervised machine learning models based on the
regression technique to solve CC equations\footnote{We did not benchmark machine learning models such as Neural Networks. On initial application, the model faced mainly two challenges. Firstly, The model requires a lot of data to train since the functional form is completely unknown. Second, the model requires a much larger amount of training time as compared to the regression model for a good enough accuracy, and the final model is inconsistent for different runs}. 
The average accuracy and computation time 
requirements for various models under consideration are
summarized in Table 1. The CC-ML(KRR) seems to be the most 
stable and robust model with guaranteed convergence 
beyond $\mu H$ at the cost of large number of training
iterations. 
The model saves around 30-40\% computation time over the
unaccelerated canonical CCSD calculations, and 20-30\%
computation time over the DIIS accelerated CCSD calculations. 
The bottleneck for the CC-ML(KRR) 
model is the high number of iterations required for the 
training on top of a few discarded initial iterations. 
The CC-ML(CKR) and 
CC-ML(PKR) perform much better in terms of the time requirement
since it can be trained with fewer number of iterations without
discarding the initial steps. For both the models, the accuracy
is marginally poorer than CC-ML(KRR) model, although the average 
error in both the cases is only a few $\mu H$ irrespective of 
the molecular correlation complexity. 
Apart from that, a relatively 'loose' classification based KNN
machine learning model gives a respectable accuracy of 9 $\mu H$
for molecules in equilibrium geometries and 40$\mu H$ in
geometries away from the equilibrium, with the time saving
similar to the previous two models.   
We thus conclude that the hybrid CC-ML technique is 
statistically stable and could be used as a standardized method
of calculation. The method works well with several supervised 
machine learning models, and is highly tunable as per the 
requirement of accuracy and cost affordability. This is also 
a demonstration of the robustness of the synergistic 
interdependence of the cluster amplitudes and the resulting
hybrid CC-ML models under various electronic complexity.  
In our implementation, the construction of 
the diagrams for only the selected excitations belonging to the 
LAS, as shown in step-II, Fig. \ref{fig:circ-caus}, is far 
from being optimal, and there is plenty of room to further 
improve upon. This would further reduce the computation time
significantly. We note that the hybrid CC-ML scheme does
not require any previously computed data; rather they 
can be trained on the fly based on the various cluster
amplitudes determined at the initial steps during the
optimization process for individual molecules.

 \section{Future Directions}
This work reinforces the synergistic interrelation of the cluster
amplitudes during the CC iteration scheme, and demonstrates the
effectiveness of the hybrid CC-ML methodology. However, being one
of the first instances of the numerical inclusion of Synergetics via ML
in CC, the possibilities are endless. Since the field of ML is
still new, development of better models that resonate with the
exact analytical structure of the CC iteration scheme would be an
exciting avenue to explore. The development of an analytical
mapping is a highly non-trivial challenge. While the adiabatic
decoupling scheme to map the auxiliary amplitudes in terms of 
the principal amplitudes is available\cite{agarawal2021adiabatic}
in literature, conversion of the technique to machine learning 
based methods for numerical efficiency is a challenge. 
Other areas of work would be the
inclusion of DIIS in the CC-ML technique for even faster
calculations. Inclusion of higher order terms like triples and 
quadruples would be an interesting avenue to venture. This would 
make high order calculations with large basis sets faster in 
near future. An extension of this model to treat
molecular excited states would also be a subject to a forthcoming
publication.
\section*{Data Availability}
The data generated in this study is available upon reasonable request to the corresponding author. 
\section*{Acknowledgement}
The authors thank Mr. Anish Chakraborty for many stimulating
discussions about the structure of the program.

\section*{Conflict of Interest}
The authors declare no competing financial interest.
\section*{Funding}
The authors thank IRCC, Indian Institute of Technology Bombay for the research seed grant, and SERB, Department of Science and Technology, Government of India for their financial support. 
\providecommand{\latin}[1]{#1}
\makeatletter
\providecommand{\doi}
  {\begingroup\let\do\@makeother\dospecials
  \catcode`\{=1 \catcode`\}=2 \doi@aux}
\providecommand{\doi@aux}[1]{\endgroup\texttt{#1}}
\makeatother
\providecommand*\mcitethebibliography{\thebibliography}
\csname @ifundefined\endcsname{endmcitethebibliography}
  {\let\endmcitethebibliography\endthebibliography}{}

\end{document}